\documentclass[twoside]{article}
\usepackage{fleqn,espcrc2}

\usepackage{graphicx}
\usepackage[figuresright]{rotating}

\def\Journal#1#2#3#4{{#1} {\bf #2}, #3 (#4)}
\def\NPB{{\em Nucl. Phys.} B}
\def\NPBPS{{\em Nucl. Phys.} B (Proc. Suppl.)}

\def\IJMPC{{\em Int. Journ. of Mod. Physics}  C}

\def\PRD{{\em Phys. Rev.} D}

\def\SS{{\em Sci. Sinica.}}
\def\Dag#1{{#1}^\dagger}
\def\eps{\epsilon}
\def\be{\beta}
\def\bek{\beta^{-k/2}}
\def\Amu{A_{\mu}}
\def\Amuk{A_{\mu}^{(k)}}
\def\Umu{U_{\mu}}
\def\Umuk{U_{\mu}^{(k)}}
\def\DLie{\nabla^{i}_{x,\mu}}
\def\Tr{\mbox{Tr}\;}
\def\Real#1{\mbox{Re\,}({#1})}
\def\beq{\begin{equation}}
\def\eeq{\end{equation}}
\def\non{\nonumber}
\def\beqn{\begin{eqnarray}}
\def\eeqn{\end{eqnarray}}
\def\frac#1#2{ {{#1} \over {#2} }}

\newcommand{\AmS}{{\protect\the\textfont2
  A\kern-.1667em\lower.5ex\hbox{M}\kern-.125emS}}

\title{Fermionic Loops in Numerical Stochastic Perturbation Theory} 

\author{F. Di Renzo\address{Dipartimento di Fisica, Universit\`a di Parma 
	and INFN, Gruppo Collegato di Parma, Italy}
        and
        L. Scorzato\address{Department of Physics, 
	University of Wales at Swansea, United Kingdom}}
       
\begin{document}

\begin{abstract}
We discuss the inclusion of fermionic loops contributions in Numerical 
Stochastic Perturbation Theory for Lattice Gauge Theories. We show how 
the algorithm implementation is in principle straightforward and report 
on the status of the project. 
\vspace{1pc}
\end{abstract}

\maketitle

\section{Introduction}
In recent years the Numerical implementation of Stochastic Perturbation 
Theory (NSPT) was introduced, which was able to reach unprecedented 
high orders in perturbative expansions in Lattice Gauge Theories (LGT). 
Till now the main limitation of the method has been the quenched 
approximation, an inclusion of fermionic loops contributions missing. 
We can now fill this gap. Let us first of all remind the basics of the 
method. \\
NSPT \cite{NSPTnew} comes (almost for free \ldots) as an application of the 
Stochastic Quantization scheme of Parisi and Wu \cite{Parisi}. We briefly 
sketch it in the context of LGT. The Cornell group showed \cite{Batrouni} 
that a Langevin equation for LGT can be formulated as
\beq\label{eq:LangEq}
\Umu(x;\tau+\eps) = e^{-F[U(\tau),\eta]} \, \Umu(x;\tau)
\eeq
where
\beqn\label{eq:F}
F[U(\tau),\eta] &=& \sum_i T^i F_i \non \\  
		&=& \sum_i T^i (\eps \DLie S[U] + \sqrt{\eps} \eta^i).
\eeqn
In the previous formulae $S[U]$ is the action for the gauge fields, $\eps$ 
is a time step and 
$\eta$ is a gaussian noise, while $\DLie$ is a Lie derivative on the 
group, whose definition can be easily understood in terms of 
\beq
f(e^{\alpha \cdot T} U) = f(U) + \alpha^i \nabla^i f(U) + O(\alpha^2).
\eeq
Eq.~(\ref{eq:LangEq}) should be understood as follows. One has at hand the 
theory described by the action $S[U]$ and the goal is to compute 
observables in terms of the functional integral, in which the measure is 
dictated by $\exp(-S[U])$ (we adhere to an euclidean formulation). In 
Eq.~(\ref{eq:LangEq}) an extra dependence on a new parameter $\tau$ is 
imposed on the fields. This parameter can be thought of as a stochastic time 
in which an evolution takes places according to Eq.~(\ref{eq:LangEq}), 
which is stochastic due to the presence of $\eta$. Now the key points are 
the gaussian nature of $\eta$ and the fact that the so called drift term 
in the $F$ appearing in Eq.~(\ref{eq:LangEq}) is given by the equations of 
motion. One can show that because of that the Langevin equation describes 
a stochastic process whose (asymptotic) equilibrium distribution is given 
just by the measure one is interested in ($\exp(-S[U])$). Therefore one can 
trade expectation values with respect to the latter for means over the 
stochastic time evolution dictated by Eq.~(\ref{eq:LangEq}). 
In the case of pure gauge Wilson action
\beq
S_G = - \frac{\be}{2N} \sum_P \Tr (U_P + \Dag{U_P})
\eeq
Eq.~(\ref{eq:F}) reads
\beq\label{eq:FSG}
\sum_i T^i \DLie S_G[U] = \frac{\be}{4N} \sum_{U_P\ni\Umu(x)} 
(U_P - \Dag{U_P})_{\mbox{tr}}
\eeq
(the subscript $\mbox{tr}$ asks for the traceless part). One can 
recognize in Eq.~(\ref{eq:FSG}) a pretty local nature, as expected. 
It is well known \cite{Parisi} that one can get a Stochastic Perturbation 
Theory from the Langevin equation. NSPT gets it by directly 
expanding the fields as 
\begin{eqnarray}\label{eq:expans}
\Umu(x) &=& \exp\left[\Amu(x)\right] \non \\
	&=& \exp \left[\sum_{k>0} \bek \Amuk(x)\right] \\
	&=& 1 + \sum_{k>0} \bek \Umuk(x). \non
\end{eqnarray}
Such an expansion has to be thought of as a formal series 
motivated by the dependence of the solution to Eq.~(\ref{eq:LangEq}) on 
the coupling ($\bek \sim g$). As a consequence every quantity depending on 
the fields can now be given a similar power expansion such as (for example) 
\beq\label{eq:Fexpans}
F[U] = \sum_{k>0} \bek F^{(k)}.
\eeq
Eq.~(\ref{eq:LangEq}) gets translated into a hierarchy of equations exactly 
truncable at any given order and suitable for a numerical integration on 
a computer. Each order of Perturbation Theory is now obtained from means 
over stochastic time evolution of suitable (maybe complicate) composite 
operators (just like those appearing in Eq.~(\ref{eq:Fexpans})). 
Notice that having decompactified 
the formulation in Eq.~(\ref{eq:expans}), divergencies in the 
non--gauge--invariant sector show up, which are cured by Stochastic Gauge 
Fixing. 

\section{Introducing fermions}
In principle introducing fermions is straightforward. One needs 
to face a new measure in the functional integral of the form
\beq
e^{-S_G} \det M = e^{-(S_G-Tr \ln M)}.
\eeq
We write $M$ for the fermion matrix to recall that other actions have the 
same structure, for example the Faddeev---Popov action \cite{Lat98}. 
One would in principle simply need to replace
\beqn
\DLie S_G &\mapsto& \DLie S_G - \DLie \Tr \ln M = \non \\
	  &       & \DLie S_G - \Tr ( (\DLie M) M^{-1}).
\eeqn
As a matter of fact, one has to face an inverse, {\em i.e.} non--locality. 
The Cornell group \cite{Batrouni} gave to the problem a solution which was 
in a sense a prototype for fermionic simulations. This amounts to rewrite 
in Eq.~(\ref{eq:F}) 
\beq\label{eq:Fnew}
F_i=\eps {\cal F}_i + \sqrt{\eps} \eta_i 
\eeq
\vspace{-0.5cm}
\beqn
{\cal F}_i = \left[\DLie S_G - 
     \Real{\Dag{\xi_k}(\DLie M)_{kl}(M^{-1})_{ln}\xi_n}\right] \non
\eeqn
in which a summation over repeated indices is to be understood. One should 
also keep in mind that $k,l,n$ are multi--indices. A new gaussian field has 
been introduced, normalized as $\langle \xi_i \xi_j \rangle = \delta_{ij}$. 
The evolution of the process will now also average over $\xi$, resulting in 
\beqn
\langle{\cal F}_i\rangle_\xi &=& \left[\DLie S_G - 
     \Tr((\DLie M)M^{-1})\right] \non \\
&=& \DLie \left[S_G - \Tr(\ln M) \right] 
\eeqn
which is exactly what one is interested in. Life is now pretty easier, since 
one simply needs to invert the fermion matrix on a source solving the system 
\beq
M_{kl} \psi_l = \xi_k
\eeq
in terms of whose solution the evolution is local:
\beq
{\cal F}_i = \left[\DLie S_G - 
     \Real{\Dag{\xi_l}(\DLie M)_{ln}\psi_n}\right]. 
\eeq
It takes a little time to realize that life is in a sense even easier in 
NSPT. To understand why, one should first of all remember that also the 
fermionic matrix gets expanded as a power series
\beq
M = M^{(0)} + \sum_{k>0} \bek M^{(k)}.
\eeq
The inverse of such a matrix is easy to compute
\beqn
M^{-1} &=& \sum_{k=0} \bek {M^{-1}}^{(k)} \non \\
       &=& {M^{(0)}}^{-1} + \sum_{k>0} \bek {M^{-1}}^{(k)}. 
\eeqn
The notation enlightens the fact ``the zeroth-order of the inverse is the 
inverse of the zeroth-order''. Other orders are not difficult to compute, 
resulting out of a simple recursion
\beqn\label{eq:rec}
{M^{-1}}^{(1)} &=& - {M^{(0)}}^{-1} M^{(1)} {M^{(0)}}^{-1} \\
{M^{-1}}^{(2)} &=& - {M^{(0)}}^{-1} M^{(2)} {M^{(0)}}^{-1} \non \\
	       && - {M^{(0)}}^{-1} M^{(1)} {M^{-1}}^{(1)} \non \\
{M^{-1}}^{(3)} &=& - {M^{(0)}}^{-1} M^{(3)} {M^{(0)}}^{-1} \non \\
	       && - {M^{(0)}}^{-1} M^{(2)} {M^{-1}}^{(1)} \non \\
	       && - {M^{(0)}}^{-1} M^{(1)} {M^{-1}}^{(2)} \non \\
\ldots		&& \non 
\eeqn
An NSPT algorithm for fermions is now easy to implement in force of the 
four considerations we now proceed to make. 
\subsection{Locality}
First of all, from the new random fields $\xi$ define the (power--expanded) 
fields 
\beq
\xi^{(j)} \equiv {M^{-1}}^{(j)} \xi
\eeq
in terms of which the $(l+j)$th contribution to Eq.~(\ref{eq:Fnew}) reads 
\beq
\xi_k \left(\DLie M \right)_{kl}^{(l)} \xi_l^{(j)} 
+ (l \leftrightarrow j, \; {\mbox if } \; l \neq j)
\eeq
The expression of $\left(\DLie M \right)_{kl}^{(l)}$ is easy to work out. 
In order not to obscure the main point with trivial algebra, we do not 
write it down, only stressing that it is as local as its non--perturbative 
counterpart, as obviously expected. 
\subsection{An easy recursive formula}
Having made the point that a simple recursive formula holds for 
Eq.~(\ref{eq:rec}), it is straightforward to notice that a similar relation 
holds for the computation of the fields $\xi^{(j)}$
\beq\label{eq:xij}
\xi^{(0)} = {M^{(0)}}^{-1} \xi 
\eeq
\vspace{-0.5cm}
\beqn
\xi^{(1)} = - {M^{(0)}}^{-1} M^{(1)} \xi^{(0)} \non 
\eeqn
\vspace{-0.5cm}
\beqn
\xi^{(2)} = - {M^{(0)}}^{-1} \left[M^{(2)} \xi^{(0)} + 
M^{(1)} \xi^{(1)} \right] \non 
\eeqn
\vspace{-0.5cm}
\beqn
\xi^{(3)} = - {M^{(0)}}^{-1} \left[M^{(3)} \xi^{(0)} + 
M^{(2)} \xi^{(1)} + M^{(1)} \xi^{(2)} \right] \non 
\eeqn
\vspace{-0.5cm}
\beqn
\ldots		 \non 
\eeqn
The message from Eq.~(\ref{eq:xij}) is quite clear: at every order only 
one proper inverse is needed and everything that is left comes out of 
a simple recursive relation in terms of already computed (local) quantities. 
On top of that, the main point still needs to be made as far as the matrix 
${M^{(0)}}^{-1}$ is concerned.
\subsection{The call for an FFT}
Being the $0$th order, ${M^{(0)}}^{-1}$ of course does not depend on the 
fields: it is actually the standard Feynmann propagator, which is 
diagonal in Fourier space. Notice that the latter observation holds 
for any fermionic matrix and also for the case of the Faddeev--Popov 
matrix \cite{Lat98}. With this respect the method is quite general, 
opening the way to multiple applications, among which Neuberger fermions, 
whose perturbative expansions have till now only been pioneered. Given 
the above observations, the obvious way to implement the construction of 
the fermionic contribution to the drift $F$ is to go back and forth from 
Fourier space. This of course calls for an efficient {\em FFT}. 
The implementation we are working on is the $4$--$d$ version of the algorithm 
described in \cite{FFT}. 
In general, one can always think of a multidimensional {\em FFT} 
as the result of subsequent applications of $1$--$d$ {\em FFT}. This approach 
is the obvious choice on an architecture such that of the {\em APE} 
family computers, in which case the algorithm basically amounts to a 
$1$--$d$ (local) {\em FFT} interlaced with transpositions\footnote{Notice by 
the way that the reason why we have differred till now the implementation of 
fermions in NSPT is just the fact that an {\em FFT} is easier on 
{\em APEmille} than on {\em APE100}}. 
Implementing an {\em FFT} in the NSPT context 
also opens the way to quite interesting byproducts. For example, one can 
think of obtaining NSPT results directly in momentum space. Also 
the possibility of exploiting the so-called Fourier acceleration can be 
taken into account. 
\subsection{Dealing with different time scales}
By inspecting Eq.~(\ref{eq:Fnew}) one in general does not expect the 
characteristic times to be the same for the pure gauge and the fermionic 
contributions. Given the overhead imposed by the inclusion of fermions it 
is important not to waste computing time. This is quite easy to do because 
in order to correct for the finite time step $\eps$ in the integration of 
the Langevin equation we adhere to the simplest recipe, that is Euler 
scheme plus extrapolation $\eps \rightarrow 0$. Now a first order 
prescription for the integration of an equation of the type 
\beq
f'(t) = g(t)+h(t)
\eeq
in which different characteristic times are present is very easy to 
work out, a trivial example being
\beqn
f(t+\eps)  &=& f(t)+ \eps g(t) \\
f(t+2\eps) &=& f(t+\eps) + \eps g(t+\eps) + 2\eps h(t+\eps). \non
\eeqn
Having said that, the building blocks representation for our algorithm 
is basically made out of three modules:
\begin{itemize}
\item
Evolution by the pure gauge contribution to the drift $F$ with a certain 
time step $\eps$.
\item
Generation of the $\xi^{(j)}$; this is the only non--local piece of the 
computation, the non--locality being anyway traded for multiple 
applications of an {\em FFT}, after which every operation is trivial. 
\item
Evolution by the fermionic contribution to the drift $F$ with a certain 
time step $\eps'$. This does not present any structural 
difference with respect to the first module. 
\end{itemize}
Of course the second and the third modules are tied together and their 
balance to the first is fixed by the considerations sketched a few lines 
above. 

\section{Conclusions, {\em i.e.} our shopping list}
As already said, the implementation of the program is in progress, so that 
our conclusions are mainly a shopping list of applications we are working on. 
\begin{itemize}
\item
The highest priority is given to the unquenched extension of the $\alpha^3$ 
computation of the Lattice Heavy Quark residual mass, which is a very 
important building block in the determination of the {\em b}-quark mass. 
The quenched result ($86.2(0.6)(1.0)$) has already made quite an impact 
\cite{mb}. Being unquenced non--perturbative simulations available, the 
unquenced extension of Perturbation Theory is compelling. 
\item
The new NSPT record is by now the tenth order computation of the 
basic plaquette \cite{10loop}, which strongly confirmed the IR Renormalon 
dominance picture. Such a picture has a clear forecast for the effect 
of the inclusion of fermions, which is fixed by the change in the 
$\beta$--function coefficients. It is worthwhile to verify this. Notice 
that as a byproduct this calls for a high loop determination of the 
critical hopping parameter, which could then be compared to the 
non--perturbative computation. 
\item
Obviously, applications could be devised for coefficients needed in various 
improvement programs. We are at the moment considering the possibility of 
extending one loop further the computation of the coefficients needed in 
the twisted--mass formulation of Lattice QCD \cite{tmQCD}.
\end{itemize}

\section*{Acknowledgments}
NSPT comes out of a Parma--Milan collaboration. The authors are indebted to 
their colleagues and friends G. Burgio, G. Marchesini, E. Onofri and 
M. Pepe for many valuable discussions.

\end{document}